\documentclass[pra,aps,amsmath,amssymb,twocolumn]{revtex4}
\usepackage{epsfig}
\usepackage{bm}

\begin{document}

\preprint{}

\newcommand{\stef}[2]{$\blacktriangleright${\sc round #1:}{\em #2}$\blacktriangleleft$}

\title{Multiple quantum NMR dynamics of spin-$\frac{1}{2}$ carrying molecules of a gas in nanopores}

\author{ S.I.Doronin}
\email{s.i.doronin@gmail.com}
\author{ A.V.Fedorova}
\email{panna@icp.ac.ru}
\author{ E.B.Fel'dman}
\email{efeldman@icp.ac.ru}
\author{ A.I.Zenchuk}
\email{azenchuk@icp.ac.ru}

\affiliation{Institute of Problems of  Chemical Physics of the Russian Academy of Sciences, Chernogolovka, Moscow Region, 142432,Russia}

\begin{abstract}

We consider the multiple quantum (MQ) NMR dynamics of a gas of spin  carrying molecules in nanocavities.
 MQ NMR dynamics is determined by the residual dipole-dipole interactions which are not averaged completely due to the molecular diffusion in nanopores.
  Since the averaged non-secular Hamiltonian describing MQ NMR dynamics depends on only one coupling constant, this Hamiltonian commutes with the square of the total spin angular momentum $\hat I^2$.
   We use the basis of common eigenstates of $\hat I^2$ and the projection of  $I$  on the external magnetic field for    investigation of MQ NMR dynamics.
    This approach allows us to study MQ NMR  dynamics in systems consisting of several hundreds of spins. The analytical approximation of the stationary profile of MQ coherences is obtained.
 The analytical expressions for MQ NMR coherence intensities   of the five-spin system in a nanopore are found.
  Numerical investigations allow us to find the dependencies of intensities of MQ coherences on their orders (the profiles of MQ coherences) in systems consisting of 600 spins and even more.
    It is shown that the stationary MQ coherence profile in the considered system is an exponential one.
    
\end{abstract}


\maketitle

\section{Introduction}
Multiple quantum (MQ) NMR dynamics is a basis of MQ  NMR spectroscopy \cite{Baum1} which, in turn, is a powerful tool to study the nuclear spin distributions in different systems (for example, in liquid crystals \cite{Baum2}, simple organic systems \cite{Baum1}, amorphous hydrogenated silicon \cite{Baum3}, etc.). 
MQ NMR has been used \cite{Lac1,Tomaselli} to investigate the size of spin clusters when the growth of MQ clusters  occurs in the process of the irradiation of the spin system on the  preparation period of the MQ NMR experiment \cite{Baum1}. 
The unique possibilities of MQ NMR to study dynamics of many-spin clusters have been recently used \cite{Krojanski,Cho} for a measurement of the decoherence rate for highly correlated spin states. 
The scaling of the decoherence rate with the number of correlated spins has been also presented \cite{Krojanski}.

The theoretical description of MQ NMR  dynamics is a very difficult task because this problem is a many-spin and multiple-quantum one.
Although the first phenomenological approach was developed together with the experimental realization of MQ NMR \cite{Baum1} the consistent quantum-mechanical theory is not developed up to now  except for one-dimensional systems.
In particular, the one-dimensional spin chain with nearest neighbor double quantum Hamiltonian \cite{Baum1} is exactly solvable and it has been shown that, starting with a thermodynamic equilibrium state, only zero and double quantum coherences are produced \cite{Feldman1}-\cite{Doronin}.
This conclusion has been confirmed experimentally for relatively small excitation times \cite{Cory,Pastawski}.
However, next-nearest couplings and other distant interactions lead to higher order coherences in MQ NMR spectra. 
These interactions are beyond the exact solvable models \cite{Feldman1}-\cite{Doronin}. Thus one should apply numerical methods
in order to take into account distant spin-spin couplings. However, even supercomputer calculations allow us to study  MQ NMR dynamics of a spin chain consisting of no more than fifteen spins \cite{Feldman3}, which is
insufficient to solve some subtle problems of MQ NMR dynamics. 
In particular, the dependence of the MQ coherence intensities on their orders (the profile  of MQ coherences \cite{Lac1}) can't be found.

Nanosize systems exhibit new possibilities for  investigations of MQ NMR dynamics. 
It is well known \cite{Baugh} that dipole-dipole interactions (DDI) of spin carrying atoms (molecules) of a gas in the non-spherical nanopores in a strong external magnetic field are not averaged completely due to the molecular diffusion \cite{Baugh,Rudavets}. 
We emphasize that the residual averaged DDI are determined by only one coupling constant which is the same for all pairs of interacting spins \cite{Baugh,Rudavets}.
As a result, the averaged non-secular two-spin/ two-quantum Hamiltonian \cite{Baum1} describing MQ NMR dynamics commutes with the operator of the square of the total spin angular momentum $\hat I^2$.
Note that eigenstates of the projection of the total spin angular momentum on the external magnetic field $I_z$ are usually used as a basis (the multiplicative basis) in order to describe MQ NMR dynamics.
Since $[\hat I^2,I_z]=0$, it is suitable to study MQ NMR dynamics in a nanopore in the basis of common eigenstates of $\hat I^2$ and $I_z$. 
Since there are no transitions with changing  $\hat I^2$ in the MQ NMR experiments the problem splits into a set of simpler problems for different values of
$\hat I^2$ and it is possible to solve the problem of the exponential growth of the Hilbert space dimension with an increase of a number of spins. 
For this reason one can investigate MQ NMR dynamics in systems consisting of several hundreds of spins and solve the problem of the profile of intensities of MQ NMR coherences.

The present paper is devoted to the investigation of MQ NMR dynamics of spins coupled by the DDI in a nanopore. The paper is organized as follows.
In Sec.II, the theory of MQ NMR dynamics of spins coupled by the DDI is developed. The numerical algorithm for MQ NMR dynamics in a nanopore is also discussed in Sec. II. 
The analytical solution for MQ NMR dynamics of a five-spin system is obtained in Sec. III. The profiles of intensities of MQ NMR coherences for the systems consisting of $200 \div 600$ spins are presented in Sec. IV. Here we give also analytical approximations of the intensity profiles with smooth curves  and compare them  with the simple physical estimations.We briefly summarize our results and discuss further perspectives in the concluding Sec. V.
\section{MQ NMR dynamics of spins in a nanopore}
 
 \begin{figure}
   \epsfig{file=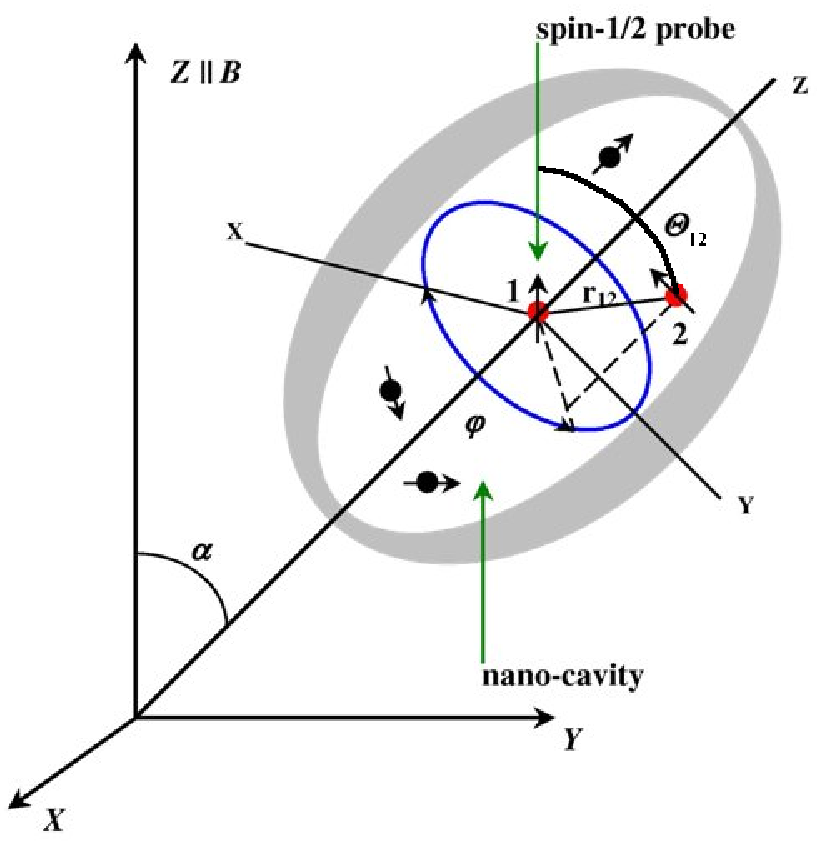, scale=0.8}
   \caption{ \label{fig:molecules} Spin-carrying molecules (atoms) in a nanopore in an external magnetic field.}
 \end{figure}

We consider a system of nuclear spins ($s=1/2$) coupled by the DDI in a strong external magnetic field in a nanopore [Fig.~\ref{fig:molecules}]. The secular part of the DDI Hamiltonian \cite{Goldman} has the following form:

\begin{equation}
\label{DDI}
H_{dz}=\sum_{j<k}D_{jk}[2I_{jz}I_{kz}-\frac{1}{2}(I_j^+ I_k^- +I_j^- I_k^+)] ,
\end{equation}
where $D_{jk}=\frac{\gamma^2 \hbar}{2 r_{jk}^3}(1-3\cos^2 \theta_{jk})$ is the coupling constant between spins $j$ and $k$, $\gamma$ is the gyromagnetic ratio, $r_{jk}$ is the distance between spins $j$ and $k$, and $\theta_{jk}$ is the angle between the internuclear vector $\vec{r}_{jk}$ and the external magnetic field  $\vec{B}$, which  is directed along z axis. The operator $I_{j\alpha}$ ($\alpha=x,y,z$) is the projection of the angular spin momentum operator on the axis $\alpha$; $I_j^+$ and $I_j^-$ are the raising and lowering operators of spin $j$.

\begin{figure}
   \epsfig{file=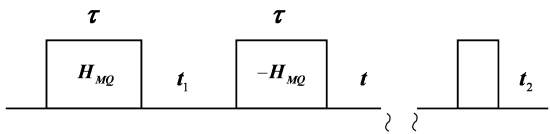, scale=0.3}
   \caption{\label{fig:scheme}The basic scheme of the MQ NMR experiment.}
 \end{figure}

The standard MQ NMR experiment consists of four distinct periods of time [Fig.~\ref{fig:scheme}]: preparation ($ \tau$), evolution ($t_1$), mixing ($\tau$) and detection ($t_2$) \cite{Baum1}. MQ coherences are created by the multipulse sequence consisting of eight-pulse cycles on the preparation period \cite{Baum1}. In the rotating reference frame  \cite{Goldman}, the averaged non-secular two-spin /two-quantum Hamiltonian, $H_{MQ}$, describing  MQ dynamics at the preparation period can be written as 
\begin{equation}
\label{RRF}
H_{MQ}=H^{(2)}+H^{(-2)},
\end{equation} 
where 
\begin{equation}
\label{Hpl}
H^{(\pm 2)}=-\frac{1}{2}\sum_{j<k}D_{jk}I_j^\pm I_k^\pm.
\end{equation}

Since the time of the molecular diffusion in nanopores is much shorter than both the dipolar time $t\approx \omega_{loc}^{-1}$ ($\omega_{loc}^2={\rm Tr}(H_{dz}^2/{\rm Tr}(I_z^2)$)\cite{Goldman} and the period of the multipulse sequence at the preparation period  of the MQ NMR experiment \cite{Baum1}, one can assume that the spin dynamics is governed by the averaged dipolar coupling constant, D, which is the same for all spin pairs. Although the problem is described by only one averaged coupling constant it is the many-spin one with the Hamiltonian 
\begin{equation}
\label{averaged}
{{\overline H}_{MQ}}=-\frac{D}{4}\{(I^+)^2+(I^-)^2\},
\end{equation}
where $I^\pm=\sum \limits_{j=1}^N I_j^\pm$ and $N$ is  the number of spins in the nanopore.

 In order to investigate MQ NMR dynamics of the system  one should find the density matrix $\rho$ solving the Liouville evolution equation \cite{Goldman} ($\hbar$=1):
   \begin{equation}
   \label{Liouville}
   i \frac{d \rho}{d \tau}=\left[{\overline H}_{MQ}, \rho(\tau) \right ]
   \end{equation}
with the initial density matrix $\rho(0)=I_z$ in the high temperature approximation \cite {Goldman}. It proves convenient to expand the spin density matrix, $\rho$, in the series as 
   \begin{equation}
   \rho(\tau)=\sum_k \rho_k(\tau),
   \end{equation}
where  $\rho_k(\tau)$ is the contribution to $\rho(\tau)$ from MQ coherences of the $k$th order. Then the intensity  $J_k(\tau)$ of the MQ coherence of order  $k$, which is an observable in the MQ NMR experiment \cite{Baum1}, is determined as
    \begin{equation}
    \label{7}
    J_k(\tau)=\frac{{\rm Tr}\{ \rho_k(\tau)\cdot \rho_{-k}(\tau) \}}{ {\rm Tr}(I_z^2)}.
    \end{equation}
The Hamiltonian  $\overline H_{MQ}$  leads to the emergence of MQ coherences of even orders only, and the coherence order cannot exceed the number of spins  $N$ \cite{Baum1}. 

Note that the eigenvalues and the eigenvectors of $\overline H_{MQ}$ have the following property. Let $\lambda$ and $u$ be the eigenvalue and the corresponding eigenvector of the Hamiltonian $\overline H_{MQ}$. Since
\begin{equation}
 e^{i \frac{\pi}{2}I_z}I^\pm e^{-i\frac{\pi}{2}I_z}=\pm i I^\pm,
\end{equation}
one can find that $-\lambda$ and $\exp (-i \frac{\pi}{2} I_z)u$ are also the eigenvalue and the eigenvector of  $\overline H_{MQ}$.

Since the square of the total spin angular momentum $\hat I^2$ commutes with projections of $I$ on  an arbitrary direction, we have from Eq.~(\ref{averaged}) that 
    \begin{equation}
    [\overline H_{MQ}, \hat I^2]=0.
    \end{equation}
Hereafter we will use the basis consisting of the common eigenstates $\hat I^2$ and $I_z$ to study MQ NMR dynamics. In this basis, the Hamiltonian $\overline H_{MQ}$ consists of the different blocks $ \overline H_{MQ}^S$ corresponding to the different values of the total spin angular momentum $S$ ($\hat{S}^2=S(S+1)$, $S=\frac{N}{2},\frac{N}{2}-1,\frac{N}{2}-2,\ldots,\frac{N}{2}-\left [\frac{N}{2} \right]$; [$i$] is an integer part of  $i$):
\begin{equation}
\overline H_{MQ}={\rm diag}\{\overline H_{MQ}^{\frac{N}{2}}, \overline H_{MQ}^{\frac{N}{2}-1}, \dots, \overline H_{MQ}^{\frac{N}{2}- \left[ \frac{N}{2} \right ]} \}.
\end{equation}
Taking into account the block structure of the Hamiltonian $\overline H_{MQ}$ and the diagonal structure of $I_z$ one can find that the density matrix $\rho (\tau )$ consists of blocks $\rho^S(\tau)$ ($S=\frac{N}{2},\frac{N}{2}-1, \dots, \frac{N}{2}-\left[ \frac{N}{2} \right ]$) as well. We will denote as $\rho_k^S(\tau)$  the contribution to $\rho^S(\tau)$ from MQ coherence of order $k$. Then the contribution $J_{k,S}(\tau)$ to the intensity of the $k$-th order MQ NMR coherence is determined as
\begin{equation}
J_{k,S}(\tau)=\frac{{\rm Tr} \left \{ \rho_k^S \cdot \rho_{-k}^S \right \}}{{\rm Tr}(I_z^2)}.
\end{equation}
 It means that the problem is reduced to the set  of the analogous tasks for each block  $\overline H_{MQ}^S$. The number of the energy levels, $n_N(S)$, for the total momentum $S$ in the $N$-spin system is  \cite{Landau}
    \begin{equation}
    n_N(S)=\frac{N!(2S+1)}{(\frac{N}{2}+S+1)!(\frac{N}{2}-S)!}, \quad 0\leq S \leq \frac{N}{2},
    \end{equation}
   which is also the multiplicity of the intensities  $J_{k,S}(\tau)$. Then the observable intensities of MQ NMR coherences $J_k(\tau)$ ($-N\leq k \leq N$) are
    \begin{equation}
    J_k(\tau)=\sum_S n_N(S) J_{k,S} (\tau).
    \end{equation}
In order to find expressions for $J_k(\tau)$ we need the matrix representations of $I^\pm$. Non-zero elements of these representations related with the total spin angular momentum  $S$  are \cite{Landau}
    \begin{eqnarray}
    \langle M|I^+|M-1 \rangle &=& \langle|M-1|I^-|M\rangle  \nonumber \\ 
     && =\sqrt{(S+M)(S-M+1)},
    \end{eqnarray}
where $M=-S+1,-S+2,\ldots,S-1,S$. Hence one can find that
       \begin{eqnarray}
       \label{elements}
      &&\langle M|(I^+)^2|M-2\rangle=\langle M-2|(I^-)^2|M\rangle  \\
      =\lefteqn{\sqrt{(S+M)(S+M-1)(S-M+1)(S-M+2)},}\nonumber    
    \end{eqnarray}   
where $M=-S+2, -S+3,\ldots,S-1,S$ and all other elements are zero. 

Eq.~(\ref{elements}) determines the matrix representation of the block $\overline H_{MQ}^S$. The dimension of this block is $2S+1$. The total dimension of the Hamiltonian, $2^N$, is determined by the sum of the dimensions of all the blocks as follows 
 \begin{eqnarray}  
  \label{dimension}
   \sum_S {n_N(S)(2S+1)} &  =&\sum_S {\frac{N!(2S+1)^2}{(\frac{N}{2}+S+1)!(\frac{N}{2}-S)!}}\nonumber\\
   &&=2^N. 
 \end{eqnarray}
The proof of the relationship (\ref{dimension}) is given in Appendix.

Now we use one more symmetry of $\overline H_{MQ}$.
Since the Hamiltonian $\overline H_{MQ}$ and all its blocks are invariant with respect to the rotation by the angle 
$\pi $ about the $z$ axis, the operator $\exp (i\pi I_z)$ is an integral of motion and
\begin{equation}
\left[ \overline H_{MQ}, \exp(i\pi I_z) \right ]=0.
\end{equation}
It means that the  $2^N\times 2^N$ Hamiltonian matrix is reduced to two $2^{N-1} \times 2^{N-1}$ submatrices. The same is valid for all blocks  $\overline H_{MQ}^S$ which split into two subblocks: 
$\overline H_{MQ}^{+S}$ and $\overline H_{MQ}^{-S}$.
 This  reduction is valid for arbitrary $N$. However for odd $N$,  both submatrices yield the same contribution into the profile of the MQ NMR coherences \cite{Feldman3}. Consequently, one should solve the problem using only  one $2^{N-1} \times 2^{N-1}$ submatrix and double the obtained intensities 
 \cite{Feldman3}. Below all calculations are performed when the number of spins is odd.
 The numerical algorithm for NMR spin dynamics with the given total spin angular momentum is based on the diagonalization of the blocks $\overline H_{MQ}^S$ ($0\leq S \leq \frac{N}{2}$) of the  Hamiltonian  $\overline H_{MQ}$ and is described in Ref. \cite{Doronin}.

  \section{The exact solution for  MQ NMR dynamics of a five-spin system in a nanopore.}
We consider a system  of $N=5$ spins coupled by the averaged DDI in a nanopore. 
The values $S$ of the total spin angular momentum are 5/2, 3/2 and 1/2 for this case.
 It was mentioned in Sec.II that if the number of spins is odd then each block $\overline H_{MQ}^S$ splits into two ones $\overline H_{MQ}^{+S}$ and $\overline H_{MQ}^{-S}$ and we can consider only one of them in order to calculate the intensities of the MQ NMR coherences ~\cite{Feldman3}.
Using Eq.~(\ref{elements}) one can obtain the eigenvalues $\lambda_{5/2}^{(i)}$ $(i=1,2,3)$ of  $\overline H_{MQ}^{+5/2}$ which are the following
\begin{equation}
 \label{eigenvals}
 \lambda_{5/2}^{(1)}=-\sqrt 7 D,\quad \lambda_{5/2}^{(2)}=\sqrt 7 D, \quad \lambda_{5/2}^{(3)}=0.
\end{equation}
The appropriate set of eigenvectors reads as follows
\begin{eqnarray}\label{eigenvectors}
   u_{5/2}^{(1)}&=&\left( \frac{1}{2} \sqrt{\frac{5}{7}},-\frac{1}{\sqrt{2}}, \frac{3}{2\sqrt 7}\right );  \nonumber\\ 
   u_{5/2}^{(2)}&=&\left(\frac{1}{2} \sqrt{\frac{5}{7}},\frac{1}{\sqrt 2},\frac{3}{2 \sqrt 7 } \right );
   \label{eigenvecs}\\
   u_{5/2}^{(3)}&=&\left(   - \frac{3}{\sqrt 14}, 0, \sqrt{\frac{5}{14}} \right )\nonumber.
\end{eqnarray}
Similarly, 
the eigenvalues and eigenvectors of the block 
$\overline H_{MQ}^{+3/2}$ are following
 \begin{eqnarray}\label{eigenvectors3}
   \label{16}
   &&
   \lambda_{3/2}^{(1)}=-\frac{\sqrt{3}}{2} D, \quad \lambda_{3/2}^{(2)}=\frac{\sqrt 3}{2} D,\\\nonumber
   &&
   u_{3/2}^{(1)}=\left( -\frac{1}{\sqrt 2}, \frac{1}{\sqrt 2} \right );
   \quad 
   u_{3/2}^{(2)}=\left (\frac{1}{\sqrt{2}},     \frac{1}{\sqrt 2} \right ).
 \end{eqnarray}
 The block $\overline H_{MQ}^{+1/2}$ is a scalar, 
 \begin{eqnarray}\label{H12}
 \overline H_{MQ}^{+1/2} =0.
  \end{eqnarray}
 The solutions of Eq.~(\ref{Liouville}), $\rho _n ^{+5/2}$ ($n=1,3,5$), where the Hamiltonian  $\overline H_{MQ}$ is changed by the Hamiltonian $\overline H_{MQ}^{n/2}$, is
 \begin{eqnarray}
   \label{rho}
   \rho^{+n/2} (\tau)&=& 
   U_{+n/2}e^{-i \Lambda ^{+n/2} \tau}U_{+n/2}^+ \nonumber\\
   &&{}\times \rho_0^{+n/2} 
   U_{+n/2} e^{i \Lambda^{+n/2} \tau}U_{+n/2}^+, 
 \end{eqnarray}
 where $\Lambda^{+n/2}$  is the diagonal matrix of eigenvalues and  $U_{+n/2}$ is the matrix of eigenvectors   of the 
   block  $\overline H_{MQ}^{+n/2}$ ($n=1,3,5$), and the initial density matrices $\rho_0^{+n/2}$ are the following
 \begin {eqnarray}\label{rho_0}
  &&
  \rho^{+1/2}_0=1/2,
  \;\;
  \rho^{+3/2}_0=
   \left(
     \begin{array}{cc}
     3/2&0\\
     0&-1/2
     \end{array}
   \right),
  \\\nonumber
  &&
  \rho^{+5/2}_0=
   \left(
     \begin{array}{ccc}
     5/2&0&0\\
     0&1/2&0\\
     0&0&-3/2
     \end{array}
   \right).
 \end{eqnarray}
After calculations using  Eqs.~(\ref{eigenvals},\ref{eigenvectors},\ref{rho},\ref{rho_0}) with $n=5$ one obtains
  \begin{equation}
  \label{genmatrice}
    \rho^{+5/2}(\tau)=
     \left(
       \begin{array}{ccc}
        a_{11}&i a_{12}&a_{13}\\
        -i a_{12}&a_{22}&i a_{23}\\
        a_{13}&-i a_{23}&a_{33}
       \end{array}
     \right),
  \end{equation}
where
  \begin{eqnarray}
   \label{matrice members}
   a_{11}&=&\frac{5}{98} \left[ 36 \cos (\sqrt 7 D \tau)- 2\cos(2\sqrt 7 D \tau)+15 \right],\nonumber\\
   a_{12}&=&\frac{1}{7} \sqrt{ \frac{10}{7}} \left[ 2 \cos(\sqrt 7 D \tau)-9 \right] \sin(\sqrt 7 D \tau),\nonumber\\
   a_{13}&=&-\frac{24}{49} \sqrt 5 \sin^4 \left( \frac{\sqrt 7 D \tau}{2} \right ),\nonumber \\
   a_{22}&=&\frac{1}{14} \left[ 4 \cos (2 \sqrt 7 D \tau)+3 \right ],\\
   a_{23}&=&-\frac{3}{7} \sqrt \frac{2}{7} \left [ 2 \cos(\sqrt 7 D \tau )+5 \right] \sin(\sqrt 7 D \tau),\nonumber\\
   a_{33}&=&-\frac{3}{98} \left[ 60 \cos(\sqrt 7 D \tau) +6 \cos (2 \sqrt 7 D \tau)-17 \right] \nonumber.
  \end{eqnarray}
 The analogous calculations for the matrix $\rho^{+3/2}$ using
 Eqs.~(\ref{eigenvectors3},\ref{rho},\ref{rho_0}) with $n=3$ yield
   \begin{eqnarray}
   \label{rho32}
   \rho^{+3/2}&=&
    \left(
     \begin{array}{cc}
      \cos(\sqrt 3 D \tau)+\frac{1}{2}& - i \sin( \sqrt 3 D \tau)\\
      i \sin(\sqrt 3 D \tau)& \frac{1}{2} - \cos (\sqrt 3 D  \tau)
     \end{array}
    \right).
   \end{eqnarray}
   Finally, Eq.(\ref{rho}) with $n=1$ in view of Eqs.(\ref{H12}) and (\ref{rho_0}) yields
   \begin{eqnarray}\label{rho12}
   \rho^{+1/2}\equiv \rho^{+1/2}_0=\frac{1}{2}.
   \end{eqnarray}
Only MQ coherences of zeroth, plus/minus second, and plus/minus fourth orders appear in the considered system.
These intensities can be calculated with Eqs.~({\ref{7}), (\ref{genmatrice}-\ref{rho12}):
 \begin{eqnarray}
 \label{J0}
  J_0(\tau)&=&
  \frac{1}{20}\Big(
  n_5(5/2)\sum_{i=1}^3|\rho^{5/2}_{ii}|^2 \\\nonumber
  &&+n_5(3/2)\sum_{i=1}^2|\rho^{3/2}_{ii}|^2 +\frac{1}{4}n_5(1/2)
  \Big)
  =\\\nonumber
  &&\frac{1}{48020}\left[27825+9604 \cdot \cos(2 \sqrt3 D \tau)\nonumber \right. \\\nonumber
  &&\lefteqn{{}+2520 \cos(\sqrt 7 D \tau)+ 7560 \cos(2\sqrt 7 D \tau)}\\
  &&\left. {}+360 \cos (3 \sqrt 7 D \tau) + 151 \cos (4 \sqrt 7 D \tau) \right ] ;\nonumber
 \end{eqnarray}
 \begin{eqnarray}
 \label{J2}
   J_{\pm2}(\tau)&=&
    \frac{1}{20}\Big(
  n_5(5/2)\sum_{i=1}^2|\rho^{5/2}_{i(i+1)}|^2 \\\nonumber
  &&+n_5(3/2)|\rho^{3/2}_{12}|^2 \Big)=
   \\\nonumber
   &&
   \frac{1}{490}\left[ 95 -49 \cos(2\sqrt 3 D \tau )\right.\\\nonumber
  &&\left.{} -45\cos(2\sqrt7 D\tau)-\cos(4\sqrt 7 D \tau)\right];\nonumber
 \end{eqnarray}
 \begin{eqnarray}
 \label{J4}
   J_{\pm4}(\tau)&=&
    \frac{1}{20}
  n_5(5/2)|\rho^{5/2}_{13}|^2=
   \\\nonumber
  &&
   \frac{144}{2401} \sin^8 \frac{\sqrt 7 D \tau}{2}.
 \end{eqnarray}
One can check that the sum of all intensities of Eqs. (\ref{J0}), (\ref{J2}), (\ref{J4}) equals~1 independently of $\tau$ according to Ref. \cite{Handy}.

\section{Numerical calculations of MQ NMR dynamics of a spin system in a nanopore}
\begin{figure*}
   \epsfig{file=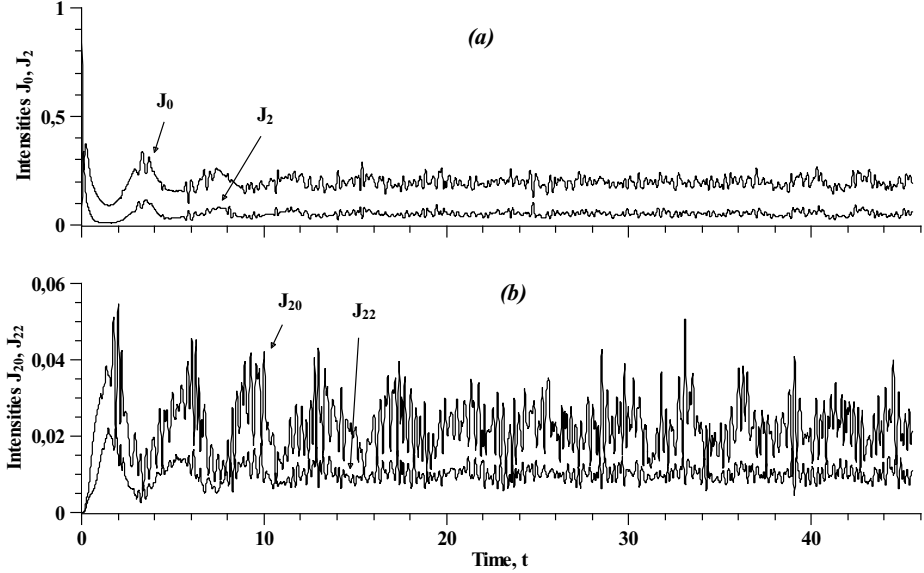, scale=0.6,angle=270}
  \caption{ Oscillating dependences of the intensities of MQ NMR coherences on the dimensionless time $t$ in the system of $N=201$ spins in a nanopore; a)$J_0$ and $J_2$; (b)$J_{20}$ and $J_{22}$. \label{fig:osc1} }
\end{figure*}
The
intensities of MQ NMR coherences are oscillating functions even at very long times.
 As an example of such behavior one can see the dependencies of intensities $J_0$, $J_2$, $J_{20}$, and $J_{22}$ of MQ coherences on the dimensionless time  $t=D\tau$ which are shown in Fig.~\ref{fig:osc1} for the system consisting of $N=201$ spins in a nanopore. 
 The minimal frequency of these oscillations is of the order $D$ and the minimal positive eigenvalue of the Hamiltonian $\overline H_{MQ}$ obtained from numerical calculations is $|\lambda _{(3/2)}^{(\min)}|=\frac{\sqrt 3}{2} D$. 
 Usually \cite{Baum1} MQ NMR dynamics is closely connected to the growth of many-spin clusters which are responsible for the emergence of MQ coherences of high orders. 
 In the beginning, the clusters consist of spins coupled by the strongest DDI which are interactions between nearest neighbors. Then, as time goes by, the clusters are growing owing to the smaller DDI corresponding to the  next nearest neighbors.  The further  growth of the clusters is determined by other remote spins.
 Thus MQ NMR coherences of high orders emerge only at long times.
 Here we have another case.
 All spins in the nanocavity are coupled through the single coupling constant $D$ and the cluster consisting of all $N$ spins emerges at the time  $t \approx 1$ ($ \tau \approx 1/D$).
 In fact the process of the growth of  many-spin clusters is absent in this system. The quasi-stationary intensity distribution develops very quickly, as it may be seen in Fig. \ref{fig:evol}, where the distributions taken at time moments $t=1,35$ almost coincide with each other. Significant discrepancies in the beginning of these graphs appear due to the oscillating behavior of the intensities.
 Nevertheless it takes some time for the emergence of the spin clusters which can absorb (emit) {\it n}-quanta ($n=0, \pm2 \pm4, \pm6$) of the energy of the external rf-fields.
 At  small times, an appearance of the corresponding terms in the density matrix $\rho$  can be clarified by the following expansion
 \begin{eqnarray}
  \rho (t)&=&e^{-i(\overline H_{MQ}/D)t}I_z e^{i(\overline H_{MQ}/D)t} \approx \\
   &&I_z-itA-\frac{t^2}{2} B+\frac{i}{6} t^3 C+ \ldots,\nonumber
 \end{eqnarray}
 where
  \begin{equation}
   A=\frac {(I^+)^2-(I^-)^2}{2},
  \end{equation}
  \begin{equation}
  B=2I_z^2-I_z-2I_z I^- I^+,
  \end{equation}
  \begin{eqnarray}
    C= - \{ 3  ( I^-)^2+9(I^+)^2-2I_z(I^-)^2-14I_z(I^+)^2 \nonumber \\
    {}+2(I^-)^3 I^+ 
    -2 I^- (I^+)^3-4 I_z^2(I^-)^2\\
     +4 I_z^2(I^+)^2\}/2. \nonumber
  \end{eqnarray}
 \begin{figure}
\epsfig{file=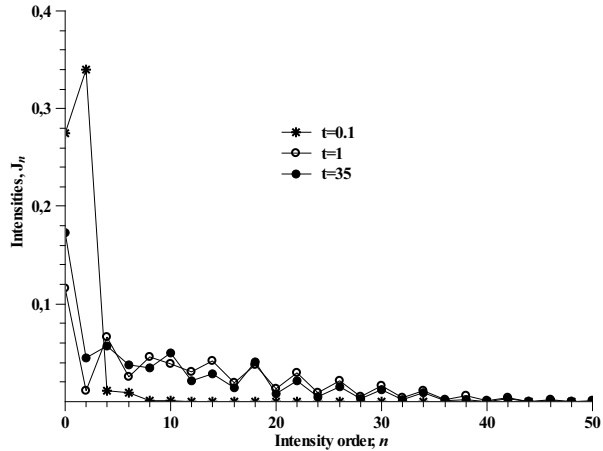, scale=0.5,angle=270}
  \caption{Evolution of the MQ NMR coherence profile in the system of $N=201$ spins in a nanopore
   at $t=0.1,1,35$.}
  \label{fig:evol}
\end{figure}

 Below we consider the intensities of MQ NMR coherences at $t\geq t_0=31$ when the coherences of all possible orders have appeared and one can think that the quasi-stationary distribution of the intensities is realized.
 We consider the averaged intensities $\overline J_k $ $(-N \leq k \leq N)$ of MQ NMR coherences with the averaging  over several periods $T= 2 \pi D/ |\lambda _{3/2}^{(\min)}|$:
   \begin{equation}
 \overline J_k=\frac{1}{K_0 T} \int\limits_{t_0}^{t_0+K_0 T} {J_k(\tau) \, d\tau},
 \end{equation}
 where $K_0$ is some positive integer taken from the requirement that the increase of the averaging interval does not change $\overline J_k $. 
 We have found  that it is  enough to take  $K_0=2$.

\begin{figure}
\epsfig{file=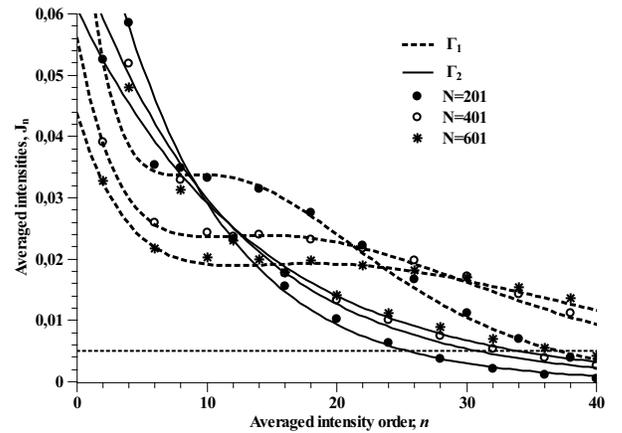, scale=0.5,angle=270}
  \caption{ The averaged intensities of MQ NMR coherences $\bar J_n$
  over the time interval $t_0 \leq t \leq t_0+2T$ with $t_0=31$ and   $T=4 \pi/ \sqrt 3$ in the systems of $N=201$, 401 and 601 spins in a nanopore.  
  }
  \label{fig:avr}
\end{figure}

 The distributions of the averaged intensities for $N=201$, 401 and 601  are shown in Fig~\ref{fig:avr}. 
 One can see that all averaged intensities may be separated into two families:
 \begin{eqnarray}
   \Gamma_1&=&\{\overline J_{4k-2}, k=1,2,\ldots\}, \\
   \Gamma_2&=&\{\overline J_{4k}, k=1,2, \ldots\}, \nonumber
 \end{eqnarray}
 while $\overline J_0$ may  not be reffered to any of them. 
 Each family may be approximated by the smooth function so that the distribution of the intensities $\overline J_{2n} $ can be represented as follows:
 \begin{equation}
  \label{distr_int}
   \overline J_{2n}=
     \left \{
     \begin{array}{r}
       A_1(1+2a_1|n|+4a_2 n^2)\exp(-2\alpha_1|n|),\\
        n=\pm1,\pm2, \ldots\\
       A_2\exp (-2\alpha_2|n|),
        n=\pm 2, \pm 4, \ldots \\
     \end{array} 
     \right..
 \end{equation}
 Parameters $A_i$, $\alpha_i$  and $a_i$ ($i=1,2$) have been found for $N=201$, 401 and 601: 
 \begin{eqnarray}
 N=201 & : & A_1=0.0875, \;\;A_2=0.0912,\\\nonumber
 &&
 \alpha_1=0,0838, \;\;\alpha_2=0.0570,\\\nonumber
 &&
 a_1=-0.0648, \;\;a_2=0.0059,\\\nonumber
 N=401 & : & A_1=0.0560, \;\;A_2=0.0704,\\\nonumber
 &&
 \alpha_1=0,0577, \;\;\alpha_2=0.0429,\\\nonumber
 &&
 a_1=-0.0433, \;\;a_2=0.0030,\\\nonumber
 N=601 & : & A_1=0.0437, \;\;A_2=0.0608,\\\nonumber
 &&
 \alpha_1=0,0465, \;\;\alpha_2=0.0367,\\\nonumber
 &&
 a_1=-0.0345, \;\;a_2=0.0020.
 \end{eqnarray}
 These parameters allow us to fulfill the normalization condition \cite{Handy}:
 \begin{equation}
 \label{norm_condition}
   \overline J_0+2\sum_{k=1}^{\infty} \overline J_k=1,
 \end{equation}
 where we take the limit $N \rightarrow \infty$.
 
 Eq. (\ref{norm_condition}) can be rewritten as 
 \begin{eqnarray}
 \label{eq31}
  \overline J_0+\frac{2 A_2}{e^{4 \alpha_2} -1}+\frac{A_1}{\sinh^3(2\alpha_1)}\left\{\sinh^2(2 \alpha_1) \right. 
  \nonumber \\
  \left .{}+a_1\sinh(4 \alpha_1) + 6a_2+2 a_2 \cosh(4\alpha_1) \right\}=1, 
 \end{eqnarray}
 where $\overline J_0$ must be found for each particular number of spins $N$. For instance, we have obtained the following three values of $\overline J_0$:
\begin{eqnarray}
\overline J_0=\left\{\begin{array}{ll}
0.1973,& N=201\cr
0.1604,& N=401\cr
0.1414,& N=601
\end{array}
\right..
\end{eqnarray}
The values of the parameters in Eq. (\ref{distr_int}) have been found according to the procedure consisting of the following three steps.
 \begin{enumerate}
  \item The family of the intensities $\Gamma_2$ of MQ NMR coherences is approximated by the function $A_2 \exp (-2 \alpha_2|n|)$ with two arbitrary parameters  $A_2$ and $\alpha_2$.
  \item The value of $A_2$ is expressed via all other parameters using Eq.~(\ref{eq31}).
  \item The family of the intensities $\Gamma_1$ of MQ NMR coherences is  approximated by the function $ A_1(1+2a_1|n|+4a_2n^2) \exp (-2 \alpha_1 |n|)$ with three arbitrary parameters $a_1$, $a_2$, and $\alpha_1$.
 \end{enumerate}
 
 The results of these approximations are represented in Fig. \ref{fig:avr}. We emphasize that
 the intensities of MQ coherences have exponential dependence on their orders. 
 This model does not confirm the  Gaussian character of the MQ NMR coherence profile \cite{Baum1}.
 On the contrary our analysis yields a similar exponential profile of MQ NMR coherences to what was found for MQ NMR coherences in solids in Ref.~\cite{Lac1}.
 In addition, the possibilities of the suggested  method allow us to find also some subtle peculiarities of the MQ NMR profile which are typical for the spin clusters in nanopores.
 
 \section{Conclusions}
 We present the model in which  MQ NMR dynamics can be analyzed with  numerical methods for systems consisting of several hundreds of spins. 
 It is essential that this model can be experimentally investigated by the methods of MQ NMR spectroscopy.
 Since the averaged coupling constant  $D$ depends on the volume and shape of the nanopore and its orientation relatively to the external magnetic field \cite{Baugh, Rudavets}, one can obtain this information from the comparison of the calculated MQ NMR spectra with  the experimental ones.
 The developed method can be also used in order to investigate MQ NMR dynamics in fluctuating nano-containers \cite{Feldman4}.
 Then the information  about correlation times of these fluctuations
 can be obtained \cite{Feldman4}.
 
Usually \cite{Baum1}, MQ NMR dynamics is studied on the basis of the averaged non-secular two-spin/two-quantum Hamiltonian of Eq.~\ref{RRF}. At the  same time, this approach allows us to take into account  the corrections due to the higher order terms of the averaged Hamiltonian theory \cite{Waugh}, since these terms commute with $\hat I^2$ as well.
  MQ NMR dynamics  can be also investigated for systems with dipolar ordered initial state  \cite{Furman,DorFeld} by the suggested method.
  
  Our approach demonstrates that, on the one hand, the MQ NMR spectroscopy is very useful tool for the investigations of the nano-size systems and, on the other hand, the study of  such systems allows us to clarify some subtle peculiarities of the MQ NMR  dynamics.
 
\begin{acknowledgments}
We are grateful to D.~E.~Feldman  for stimulating discussions.
The work was supported by the Program of the Presidium of RAS No.~27 "Foundations of fundamental investigations of nanotechnologies and nanomaterials".
\end{acknowledgments}

\appendix* 

\section{}
In order to prove the equality~(\ref{dimension}) we use the following relationships:
 \begin{eqnarray}
 \label{A1}
  \sum_{S=0}^{N/2}{C_{N+1}^S}&=&2^N; 
 \sum_{S=1}^  {N/2} {C_N^{S-1}}=2^{N-1}-\frac{1}{2}C_N^{N/2}\nonumber; \\
  &&\sum_{S=2}^{N/2} {C_{N-1}^{S-2}=2^{N-2}}-\frac{1}{2}C_N^{N/2},
 \end{eqnarray}
where $C_N^S={N \choose S}=\frac{N!}{S!(N-S)!}$. The simple transformations yield
 \begin{eqnarray}
 \label{A2}
   \sum_{S=0}^{N/2}{\frac {N!(2 S+1)^2}{(\frac {N}{2}+S+1)!(\frac{N}{2}-S)!}}\\=
  \frac{4}{N+1}\sum_{S=0}^{N/2}{S^2 \cdot C_{N+1}^{N/2-S}} 
  &+\frac{4}{N+1} \sum\limits_{S=0}^{N/2} {S\cdot C_{N+1}^{N/2 -S}}\nonumber\\
   &{}+\frac{1}{N+1} \sum\limits_{S=0}^{N/2} {C_{N+1}^{N/2-S}}.\nonumber
 \end{eqnarray}
 One can find directly with Eq.(A1) that
 \begin{equation}
 \label{A3}
  \frac{4}{N+1} \sum_{S=0}^{N/2} {S C_{N+1}^{N/2 - S}}=\frac{N \cdot 2^{N+1}}{N+1}-2^{N+1}+2 C_N^{N/2},
 \end{equation}
\begin{equation}
\label{A4}
  \frac{4}{N+1} \sum_{S=0}^{N/2} {S^2 C_{N+1}^{N/2 - S}}=\frac{2^N \cdot N^2}{N+1}-(N-2)\cdot 2^N-2 C_N^{N/2},
 \end{equation}
Using Eqs.~(\ref{A1}), (\ref{A3}), (\ref{A4}) one obtains
 \begin{equation}
 \sum_{S=0}^{N/2}{\frac{N!(2S+1)^2}{(\frac{N}{2}+S+1)!(\frac{N}{2}-S)!}}=2^N
 \end{equation}
The equality of Eq.~(\ref{dimension}) is proved.

\end{document}